\documentclass[12pt,a4paper]{article}
\usepackage[utf8x]{inputenc}
\usepackage{ucs}
\usepackage[english]{babel}
\usepackage{amsmath}
\usepackage{amsfonts}
\usepackage{amssymb}
\usepackage{graphicx}
\date{}

\newcommand{\beq}{\begin{equation}}
\newcommand{\eeq}{\end{equation}}
\newcommand{\R}{\mathbb{R}}
\newcommand{\C}{\mathbb{C}}
\newcommand{\Z}{\mathbb{Z}}

\newcommand{\dd}{d}
\newcommand{\del}{\partial}
\newcommand{\D}{\mathcal{D}}
\newcommand{\nn}{\nonumber}

\title{Topology and Quantum States: the Electron-Monopole System}
\author{F.Di Cosmo, G.Marmo, A.Zampini}

\begin{document}
\maketitle
\begin{abstract}
This paper starts by describing the dynamics of the electron-monopole system at both classical and quantum level  by a suitable reduction procedure. This suggests, in order to realise the space of states for quantum systems which are classically described on topologically non trivial configuration spaces, to consider Hilbert spaces of exterior differential forms. Among the advantages of this formulation, we present in the case of the group ${\rm SU}(2)$, how it is possible to obtain all unitary irreducible representations on such a Hilbert space, and  how it is possible to write scalar Dirac type operators, following an idea by K\"ahler.  
\end{abstract}

\section{Introduction}
Dirac picture of Quantum Mechanics requires a Hilbert space of states upon which the operators associated to observables act. Even if this abstract description is very elegant and effective, it does not allow to distinguish between different physical systems. Therefore to concretely represent a quantum dynamical system one has to choose a realization of a Hilbert space. For instance the description of a particle moving on a line is achieved by adopting the Hilbert space $\mathcal{L}^2(\mathbb{R},dx)$ of square-integrable functions on $\mathbb{R}$ with respect to the Lebesgue measure. The square modulus $|\psi(x,t)|^2$ of such a wave-function is interpreted as the probability density of finding the quantum particle at the point $x$ at the time $t$: according to the physical interpretation, only the module of the wave-function must be a continuous function. This freedom may allow  to replace wave-functions by wave-sections of line bundles associated to a ${\rm U}(1)$-principal bundle, or more generally, sections of any vector bundle \cite{balga}. The first non trivial setting in which this idea was analyzed  is  the dynamics of the  electron-monopole system and one of the results was the quantization of the electric charge as suggested  by Dirac \cite{Dirac} and then more geometrically formulated  by Wu and Yang \cite{Wu-Yang}. In this paper we start by describing the dynamics  of the electron-monopole system inspired by \cite{Bal1}-\cite{Bal3}. The analysis of this system gives us the opportunity to talk about another generalization of the spaces of states, introducing the Hilbert space of square-integrable differential forms. This idea seems very attractive because differential forms contain more direct information on the topology of a manifold with respect to functions, which are also included in this new Hilbert space. 

The electron-monopole system has been widely studied during the last century because it is one of the easiest example of a dynamical system on a manifold with a non trivial topology. The configuration space of this system, in fact, is $\mathbb{R}_0^3 = \mathbb{R}^3-\left\lbrace 0 \right\rbrace$. The differential form $F$ describing the magnetic field generated by the monopole is closed but not exact and therefore it does not admit a globally defined vector potential. This fact has interesting consequences especially in the quantum formulation of the dynamics, since  the Schr\"{o}dinger equation of a particle in the magnetic field involves the vector potential.

The classical equations of motion of a charged particle in the magnetic field generated by a monopole were  introduced by Poincaré in order to explain the results of an experiment made by Birkeland, who had discoveried that cathod rays focused to a point when passed in the vicinity of a long magnet \cite{Poincare}. 
The quantum version of this dynamical system was investigated by Dirac, who  introduced a string singularity \cite{Dirac}. Even if this singularity spoiled the rotational symmetry of the problem, Dirac's work focused physicists' attention on topology because he obtained a quantization condition for the electric charge. 

As already mentioned, the work by Dirac was developed  by Wu and Yang. In order to preserve  the symmetry of the system, they  replaced wave-functions by wave-sections of a line bundle \cite{Wu-Yang}. One covers the configuration manifold with charts and in each region one can write a Schr\"{o}dinger equation; in the overlapping region  the solutions differ by a phase. In this way the modulus is a well defined continuous function, preserving the probabilistic interpretation of quantum mechanics. 

A different geometric approach to the problem is taken in \cite{Bal1}-\cite{Bal3}: to avoid the topological obstruction to the existence of a potential one enlarges  the configuration space to a ${\rm U}(1)$-principal bundle. It  is possible to define a global potential on this new manifold because on it the second cohomology class is trivial. The equations of the motion of the electron-monopole system are then obtained by introducing a Lagrangian where the added degree of freedom is not dynamical.

In this paper we exploit  the same idea,   extending the configuration space in order to evade the topological obstruction. However we derive the equations of the motion from a ``free'' Lagrangian associated to an invariant metric on the new manifold. The Hamiltonian operator turns out to be the Laplace-Beltrami operator associated to this metric. Writing the Hamiltonian operator in terms of a scalar differential operator is a starting point to discuss about the introduction of the Hilbert space of differential forms, because the Laplace-Beltrami operator can act upon  the whole exterior algebra of a manifold. 

The idea of enlarging the configuration space traces back to the work by Hertz \cite{Hertz}. He advocated the extension of the space to describe any system moving in a field of forces coming from a potential as a projection of a geodesical motion in the bigger space associated with a suitable metric tensor depending on the potential.
Later on a similar proposal was made by Kaluza and indipendently by Klein in order to incorporate also Lorentz-type forces \cite{kaluza-Klein}. Their proposal was motivated by the attempt to describe a unification of gravitation and electrodynamics in geometrical terms on a larger manifold.

Another kind of extension has been described by Duval et al.\cite{Duval} to show that the Newton-Cartan theory could be described on a five dimensional manifold with a Lorentz-type metric. This point of view allowed to describe not only the motion of the classical test particle but also the quantum motion provided by a Schr\"{o}dinger equation written in terms of a Laplace-Beltrami operator on the larger space.
In   \cite{Lizzi} again an extension of the carrier space of wave-type equations was proposed to transform the relevant differential operator into an homogeneous one, so as to avoid that the principal symbol of the operator would not contain information on the potential and on the time derivative. Again the proposal amounts to an enlargement of the configuration space to a ${\rm U}(1)$-principal bundle. If the principal symbol is also not degenerate, one may interprete it as a metric tensor and therefore giving rise to geodetical motions on an enlarged space. Solutions of the original motion are recovered by projecting the geodetical trajectories onto the original configuration space.
In  \cite{Grabowski}  the same proposal was used to deal with the transformation properties of the wave function under Galilei transformations. 

These various proposals can be grouped under the quest for a purely geometric description of the motion as the reduction of a geodetical motion on some higher dimensional Riemannian manifold. We take our point of view by exploiting the possibility of extending the set of square integrable functions by considering  Hilbert spaces of square integrable differential forms. An immediate benefit of this enlargement is the possibility of the introduction of the square root of the Laplacian, providing a description of Dirac-type operators as scalar differential operators acting on differential forms. The reduction to the usual treatment of the Dirac operator will arise from the requirement of irreducibility which are characteristic of the description of elementary particles. In summary this paper should be considered as a first attempt to describe non trivial quantum situations in terms of differential forms.

The paper is divided in two sections. The first section is devoted to the revisitation of the electron-monopole system: after recalling the  approach by \cite{Bal1, Bal2, Bal3} we present a description of the system by means of a reduction procedure from a ``free'' system on a bigger space. The Hamiltonian operator describing the ``free'' motion on a Riemannian manifold is the Laplace-Beltrami operator associated to a metric tensor on the enlarged manifold. 

The second parte deals with the proposal of replacing the Hilbert space of square integrable functions on a manifold by the Hilbert space of square integrable differential forms. Then we present two possible applications of such a generalization: one is related to representation theory whereas the other one is linked to spin geometry.

\section{Electron-monopole system}
In this section we will recall the formulation presented in \cite{Bal1, Bal2,Bal3}, which provides a global description for the dynamics of the electron - monopole system at both classical and quantum level. It is well known that, since the classical configuration space for the system -- which is  $\R_0^3\,\simeq\,R_+\times{\rm S}^2$ -- has a non trivial second homology group, the magnetic field generated by a monopole cannot be described by a globally defined potential.  In order to write a global Lagrangian and define canonical variables for the quantization, one can enlarge the configuration space to a $U(1)$ principal 
bundle over $\mathbb{R}_0^3$. 

One considers $P\,=\,\mathbb{R}_0^4 \simeq {\rm S}^3 \times \mathbb{R}_+$ as a total bundle space over the base space $\R^3_0$.  Since the  sphere ${\rm S}^3$ coincides with the  manifold of the Lie group ${\rm SU}(2)$,  we can parametrize the space $P$ by the pair $(r,s)$ with $\R\,\ni\,r\,>\,0$ and  $s$ given by  the matrix
$$
s= \left(  
\begin{array}{cc}
u & -\bar{v} \\
v & \bar{u}
\end{array}
\right)
$$   
provided  $|u|^2+|v|^2=1$. Adopting the notation $0\,<\,r^2\,=x_ix_i$ and $\hat{x}_i\,=\,x_i/r$ in $\R_0^3$,
the projection map  $\pi : \: \mathbb{R}_+ \times {\rm S}^3\: \rightarrow \: \mathbb{R}_+ \times {\rm S}^2$  defining the bundle  is given by
\begin{equation}
\mathbb{R}_+\times {\rm S}^2 \ni (\rho , \hat{x}) = (\rho , \hat{x}_i\sigma^i) = (r , s\sigma^3s^{-1})
\label{pro}
\end{equation}
where $\sigma^i$ is the $i^{th}$ Pauli matrix. 
From the following Lagrangian function  $\mathcal{L}\,\in\,\mathcal{F}({\rm T}\mathbb{R}_0^4)$, 
\begin{equation}
\mathcal{L}=\dfrac{1}{2}m \dot{r}^2+ \dfrac{1}{4}mr^2 \,{\rm Tr}\,\dot{\hat{x}}^2+ in\,{\rm Tr}(\sigma_3s^{-1}\dot{s}),
\label{Lag}
\end{equation}
the Euler-Lagrange equations in implicit form are 
\begin{align}
\ddot{r}&=r\sum_i(\dot{\hat{x}}_i)^2 &\\
\dfrac{d}{dt}\left\lbrace -\dfrac{i}{2}\left[ \hat{x},mr\dot{\hat{x}} \right] + n\hat{x} \right\rbrace &= \sum_k \dfrac{d}{dt} \left[ \epsilon_{jlk}mr^2\dot{\hat{x}}^l \hat{x}^j + n \hat{x}^k \right]\sigma^k   = \dfrac{d}{dt} (L_k \sigma^k) = 0.
\end{align}
Notice that  second equation above shows that an angular momentum -- which differs from the one corresponding to a free particle dynamics by a helicity term -- is conserved. 

The Lagrangian \eqref{Lag} admits a gauge invariance, namely it changes by a total time derivative upon the transformation 
$$
s(\hat{x})\quad\mapsto\quad s(\hat{x})\,e^{i\sigma_3\theta (t)/2},
$$   
which is given by the right action of the group ${\rm U}(1)$ upon $\R_0^4$. Such a gauge invariance provides the primary constraint -- within the formalism introduced by Dirac \cite{Dirac2} --  given by
\begin{equation}
\hat{x}_kL^k = n.
\label{con}
\end{equation}

\subsection{Canonical Quantization}
In this section  we are going to review how the electron-monopole system can be studied from a quantum point of view. The presence of constraints can be handled according to the procedure introduced by Dirac in \cite{Dirac2}. There is only the following  primary constraint
\begin{equation}
\phi \: : \: \hat{x}_a L^a - n \thickapprox 0,
\label{constr}
\end{equation}
and therefore it is also first class. The Hamiltonian for the classical dynamics is given by 
\begin{equation}
\mathcal{F}({\rm T^*}\R^4_0)\,\ni\,H= \dfrac{p_r^2}{2m}+\dfrac{(L_aL^a-n^2)}{2mr^2}+\lambda \phi
\label{Hamiltonian}
\end{equation}  
with $\lambda$  a Lagrange multiplier. 

Since we have enlarged the configuration space for the classical dynamics of the system  we consider, in order to realize  the quantum states, the Hilbert space 
\begin{equation}
\mathcal{H}\,=\,\mathcal{L}^2(\R^4_0, d\mu), \qquad\qquad d\mu\,=\,r^2\,dr\,d\nu 
\label{acca4}
\end{equation}
where $d\nu$ is the Haar measure on ${\rm S}^3\,\simeq\,{\rm SU}(2)$. Notice that the measure $d\mu$ differs from the usual $d\tilde\mu\,=\,r^3\,dr\,d\nu$ which is the restriction to $\R^4_0$ of the euclidean one on 
$\R^4$. In this way, if $d\mu^{\prime}\,=\,r^2\,dr\,d\Omega$ with $d\Omega$ the standard euclidean measure on ${\rm S}^2$, the pullback to $\mathcal{F}(\R^4_0)$ of the elements in 
$ \mathcal{H}^{\prime}\,=\,\mathcal{L}^2(\R_0^3, d\mu^{\prime})$ are elements in $\mathcal{H}$. Moreover, self adjoint operators on $\mathcal{H}$ are projectable on $\mathcal{H}^{\prime}$: by projectable we mean that, given a selfadjoint operator $T$ on $\mathcal{H}$, it is still selfadjoint on the subspace of $\mathcal{H}$ again given as the pullback to $\R^4_0$ of the elements in $\mathcal{H}^{\prime}$.

One can implement the constraint $\eqref{constr}$ by selecting a subspace $\mathcal{H}_n \subset \mathcal{H}$ 
\begin{equation}
\mathcal{H}_n = \left\lbrace \psi(r,s) \in \mathcal{H} : \hat{x}^a \hat{L}_a \psi(r,s) - n \psi(r,s)=0 \right\rbrace,
\end{equation}
where the  operators $\hat{L}_a$'s are realized as the first order  differential operators on $\mathcal{H}$ giving the right invariant vector fields corresponding to the left regular action of the Lie group ${\rm SU}(2)$ on $\mathcal{H}$.  
We associate to the Hamiltonian function \eqref{Hamiltonian} the differential operator
\begin{equation}
\hat{H}= \dfrac{1}{2mr^2}\dfrac{\partial}{\partial r}\left( r^2\dfrac{\partial}{\partial r} \right)+ \dfrac{1}{2mr^2}\left(\hat{L}_a \hat{L}^a -n^2\right) + \lambda (\hat{x}_a\hat{L}^a - n)
\end{equation}
The restriction of the action of such  Hamiltonian operator to functions $\psi \in \mathcal{H}_n$ gives the following eigenvalue equation 
\begin{equation}
\hat H\psi=\dfrac{1}{2mr^2}\dfrac{\partial}{\partial r}\left( r^2\dfrac{\partial}{\partial r}\psi \right)+ \dfrac{1}{2mr^2}\left(\hat L_a\hat L^a -n^2\right)\psi=E\psi.
\label{qHam}
\end{equation}
The solutions of this equation are well-known. They can be factorized as the product of two functions 
\begin{eqnarray*}
\psi(r,s)=R_j^n(r)D^{j}_{nm}(s)  \\
R_j^n \in \mathcal{L}^2(\mathbb{R}_+ , r^2dr) \\
D^j_{nm} \in \mathcal{L}^2({\rm S}^3 , d\nu), 
\end{eqnarray*}
where $D^{j}_{nm}$ are the Wigner functions giving a basis -- following the Peter -Weyl theorem -- for
$\mathcal{L}^2({\rm S}^3 , d\nu)$. The different indices label the action of the commuting set of operators
$\left\lbrace L^2,L_z,\hat{x}^aL_a  \right\rbrace$, that is they satisfy the equations
\begin{equation}
\begin{split}
L_z D^j_{nm} = m D_{nm}^j \\
\hat{x}^aL_a D^j_{nm} = n D^j_{nm} \\
L^2 D^j_{nm}=- j(j+1)D^j_{nm} 
\label{mon-har}
\end{split}
\end{equation}
with $j\,=\,1/2, 1, 3/2, \ldots$ labelling the irreducible representations and $n,m\,=\,-j,-j+1 \ldots,j-1 , j$.  These functions are also called monopole harmonics and are homogeneous polynomials of degree $2j$ in the variables $\left\lbrace u,v, \bar{u}, \bar{v} \right\rbrace$ (one can refer to \cite{Dray} or \cite{Schwinger rep} for a detailed computation of these polynomials).
As far as the radial part of \eqref{qHam} is concerned one has to solve the differential equation 
$$
\left[ -\dfrac{1}{2mr^2}\dfrac{\partial}{\partial r}\left( r^2\dfrac{\partial}{\partial r} \right)+ \dfrac{l(l+1)-n^2}{2mr^2} - E \right] R_l^n(r)=0
$$ 
To remove the first order term one performs the following transformation
$$
R_l^n(r)=\dfrac{u_l^n(r)}{r}
$$
The new function $u_l^n(r)$ satisfies the equation
\begin{equation}
\left[ -\dfrac{1}{2}\dfrac{\partial^2}{\partial r^2}+\dfrac{l(l+1)-n^2}{2r^2}-E\right]u_l^n(r)=0
\label{radmot}
\end{equation}
with the self-adjointness conditions
 \begin{eqnarray*}
&\int_0^{\infty}|u_l^n(r)|^2dr<\infty &\\
&\lim_{r\rightarrow 0^+}r^{-\frac{1-\sqrt{(2l+1)^2-n^2}}{2}}u_l^n(r)=1
\end{eqnarray*}
One can compare such a radial equation with the one coming from the eigenvalue problem for the quantum free particle in $\R^3$, which reads:
\begin{eqnarray*}
\left[ -\dfrac{1}{2}\dfrac{\partial^2}{\partial r^2}+\dfrac{l(l+1)}{2r^2}-E\right]u_l^n(r)=0 &\\
\int_0^{\infty}|u_l^n(r)|^2dr<\infty &\\
\lim_{r\rightarrow 0^+}r^{-l-1}u_l^n(r)=1.
\end{eqnarray*}
These relations show that the electron - monopole Hamiltonian and the free particle Hamiltonian do not share a common dense domain of self adjointness. Therefore it is not possible to treat the monopole interaction as a perturbation of the free dynamics and a partial wave analysis cannot be performed on the spherical harmonics basis. A possible solution to this problem is, indeed, the definition of the monopole harmonics \eqref{mon-har}.\newline
Coming back to the equation \eqref{radmot}, the solution of this equation for $E>0$ is a Bessel function
\begin{equation}
R_l^n(r)=\dfrac{1}{\sqrt{kr}}J_{\mu}(r)
\label{radsol}
\end{equation}
where
\begin{align*}
&\mu = \sqrt{l(l+1)-n^2+\dfrac{1}{4}}=\sqrt{\left( l+\dfrac{1}{2} \right)^2-n^2}>0 &\\
&k=\sqrt{2mE}
\end{align*}
whereas when $E<0$ \eqref{radmot} there are no meaningful solutions \cite{Wu-Yang}.

\subsection{A geometric Hamiltonian operator}
As already mentioned, the idea of describing interacting systems as a suitable reduction of free ones formulated in higher dimensional spaces has been widely developed\footnote{In particular Kaluza-Klein theories describe electromagnetic field on a 4-dimensional manifold as the curvature of a suitable  metrics on a 5-dimensional one $\cite{Bergmann}$.}.
Following this idea, 
 we shall  show in this section how the previous solutions of the electron-monopole dynamical system can be obtained in terms of a suitable reduction procedure starting from a geodesical dynamics on the enlarged manifold $\mathbb{R}_0^4$.
 
Let us consider the bundle space $\mathbb{R}_0^4$ equipped with the following metric tensor
\begin{equation}
g=dr \otimes dr + r^2 \left( \theta^1\otimes\theta^1 + \theta^2\otimes\theta^2 \right)+ k \theta^3\otimes\theta^3
\label{metr}
\end{equation}     
where $\theta^a, \, (a\,=\,1,2,3)$ are the left invariant 1-forms, dual to the left invariant derivations $X_a$ for ${\rm SU}(2)$. It is given by the superposition of  the pull-back of the metric on the base manifold $\R^3_0$ and a metric on the fibre \cite{Marmo3}, while $k$ is a numerical constant, which might indeed be given by  the pull-back of a function on $\mathbb{R}_0^3$. 
The corresponding volume form is 
$$
\Omega = \sqrt{k}r^2dr\wedge\theta^1\wedge\theta^2\wedge\theta^3
$$
A quantum description of this system is provided by considering the  Hamiltonian operator as given by the   Laplace-Beltrami operator acting on the Hilbert space $\mathcal{H} = \mathcal{L}^2(\mathbb{R}_0^4,d\mu)$ of square integrable functions on $\mathbb{R}_0^4$ with respect to the measure $d\mu = \sqrt{k}r^2drd\nu$, where $d\nu$ is the Haar measure on the sphere $S^3$. One has
\footnote{ 
In order to write the action of the Laplace-Beltrami operator on functions $\psi \in \mathcal{C}^{\infty}(M)$ on a manifold $(M, g)$ equipped with a metric tensor $g$, one introduces the gradient operator via the implicit formula  
\begin{eqnarray*}
{\rm grad} : \mathcal{C}^{\infty}(M) \: \rightarrow \mathfrak{X}(M) \\
g({\rm grad}\psi, X)\,=\,d\psi(X) 
\end{eqnarray*}
for any $X\,\in\,\mathfrak{X}(M)$ (vector fields on $M$). Afterwards one defines the 
divergence operator via again an implicit relation, namely
\begin{equation*}
L_X \Omega = ({\rm div} \, X ) \Omega
\end{equation*} 
where $X\in \chi(M)$ and $\Omega$ is the volume form on $M$ coming from the metric tensor $g$.}
\begin{align}
\Delta \psi ={\rm div} \, {\rm grad }\: \psi& = \left[ \dfrac{\partial^2}{\partial r^2}+\dfrac{1}{r^2}\left( X_1^2+X_2^2 \right)+\dfrac{1}{k}X_3^2 \right]\psi 
\label{deltag}\\ 
& \qquad = \left[ \dfrac{\partial^2}{\partial r^2}+\dfrac{1}{r^2}\left( X_1^2+X_2^2 +X_3^2 \right)-\dfrac{X^2_3}{r^2}+\dfrac{1}{k}X_3^2 \right]\psi
\nonumber 
\end{align}
where $X_a$  are the left invariant vector fields.

In order to derive the equations of motion of the electron-monopole system we implement a reduction procedure suggested by the fact that  wave functions  with different phase factors realize the same  state for a quantum system. This means that quantum states are given by the quotient of the set of wave functions on the configuration space of the system with respect to a suitable  action of a ${\rm U}(1)$ group. This notion can be naturally described in the language of principal bundles. If $\pi\,:\,P\,\to\,B$ is a principal bundle with  gauge group $G$, for  any representation $\rho\,:\,G\,\to\,{\rm Aut}(V)$ of $G$ on a vector space $V$, a   function $\psi\,:\,P\,\to\,V$ is called equivariant with respect to $\rho$ if
\begin{equation}
\psi(p\,\gamma)\,=\,\rho(\gamma^{-1})\psi(p)
\label{eq}
\end{equation}
where $p\,\in\,P$ and $p\,\gamma$ gives the right action   of the element $\gamma\,\in\,G$ upon $p$. 

One sees immediately that the operator $\Delta$ \eqref{deltag}  acts upon elements in $\mathcal{H}$, i.e. wave functions defined on the total space $\R_0^4$ of a ${\rm U}(1)$-principal bundle.   The idea now is to 
reduce the action of $\Delta$ to the subspace of square integrable $\mathbb{C}$-valued functions which are equivariant with respect to the unitary irreducible representation 
$$\rho(\varphi)=e^{in\varphi}$$
with $n\,\in\,\Z$ and $\varphi\,\in\,[0, 2\pi)$ giving an element in ${\rm U}(1)$. 
There is an interesting advantage in adopting this kind of description:  these equivariant functions are globally defined on a manifold which is parallelizable, so differential operators come from a globally defined differential calculus. 

Let us denote by $\mathcal{H}_n$ the set of square integrable equivariant functions with respect to the previous measure. From $\eqref{eq}$ we have that
\begin{equation}
\mathcal{H}_n\,=\,\{\psi\,\in\,\mathcal{H}\,:\,X_3 \psi - i n \psi=0\}.
\label{eq2}
\end{equation}
This condition is  equivalent to the constraint $\eqref{constr}$ because 
\begin{equation}
e^{-i\varphi\hat{x}_j\,\sigma^j/2}\,s\,=\,s\,e^{-i\varphi\,\sigma^3\,2}.
\label{comm}
\end{equation}
The Laplace-Beltrami operator preserves this subspace because it commutes with the vector field $X_3$. Therefore one can reduce the dynamics to $\mathcal{H}_n$. The resulting Hamiltonian operator is 
\begin{equation}
\Delta \psi = \left[ \dfrac{\partial^2}{\partial r^2}+\dfrac{L^2}{r^2}-n^2\left( \dfrac{1}{r^2}-\dfrac{1}{k} \right) \right]\psi
\end{equation}
The spectral properties of the operator above resemble those of the operator studied in the previous section. Eigenfunctions are the same, eigenvalues are shifted by a constant term.

This example shows how peculiar topological properties of a configuration space $Q$ for a system  can be taken into account by suitably extending it to the total space of a bundle.  Another possibility of taking into account the topology of a configuration space $Q$ could be that of defining the dynamics on a set of states 
related to the differential forms on $Q$, since the exterior algebra over a manifold brings brings informations on the topology of $Q$.

\section{The Hilbert space of differential forms}

In this section we consider the Hilbert space of square integrable differential forms on a manifold $M$, thus generalizing the usual Hilbert space of square integrable functions. Such an extension is suggested by the fact that the quantum Hamiltonian acts upon functions as the Laplace-Beltrami operator, and this action can be meaningfully extended upon differential forms. 

We want to briefly recall that such an extension may be of interest for many systems whose evolution is not directly governed by a Laplace-Beltrami operator. When the evolution of a system is given by an inhomogeneous second order  differential operator with terms of degree one and zero, one may transform it 
 \cite{Lizzi} into an homogeneous one  by adding a new degree of freedom. The configuration space is enlarged to a $U(1)$-principal bundle and the reduction to the original situation is achieved by considering the subspace of equivariant functions with respect to the $U(1)$-action on the new manifold. If the principal symbol associated to this new operator is not degenerate it can be interpreted as a scalar product with respect to a metric tensor. Having a metric and the corresponding metric volume one can build the relative Laplace-Beltrami operator which will describe the free motion on this space.
Let us consider the following example. If one thinks of the Schr\"{o}dinger equation written as a differential relation on the Hilbert space of square integrable functions on a manifold, possible inhomogeneous terms are due to the time derivative and to the presence of potentials or magnetic fields. As an example let us consider the differential operator 
$$
D = \left( i\dfrac{\partial}{\partial t} + \dfrac{\partial ^2}{\partial x^2} +\dfrac{\partial ^2}{\partial y^2}+ \dfrac{\partial ^2}{\partial z^2} - V(x,y,z) \right) 
$$     
By using an additional degree of freedom and the infinitesimal generator of the circle group we transform the original differential relation $D\psi = 0 $ into an homogeneous one $D^{\prime} \psi = 0$ with $D^{\prime} $ being
$$
D^{\prime} = \left( \dfrac{\partial}{\partial s}\dfrac{\partial}{\partial t} + \dfrac{\partial ^2}{\partial x^2} +\dfrac{\partial ^2}{\partial y^2}+ \dfrac{\partial ^2}{\partial z^2} + V(x,y,z)\dfrac{\partial ^2}{\partial s^2} \right)
$$ 
where $s$ is the parameter along the fibre $U(1)$. We would recover previous operator on the subspace of functions having the form
$$
\psi^{\prime}=e^{-is}\psi (x,y,z,t)
$$
The principal symbol $\cite{Carinena}$ of this differential operator is 
$$
\sigma_D = V \dfrac{\partial}{\partial s}\otimes \dfrac{\partial}{\partial s} + \dfrac{1}{2}\left( \dfrac{\partial}{\partial s}\otimes \dfrac{\partial}{\partial t} + \dfrac{\partial}{\partial t} \otimes \dfrac{\partial}{\partial s}\right) + \dfrac{\partial}{\partial x}\otimes \dfrac{\partial}{\partial x} + \dfrac{\partial}{\partial y}\otimes \dfrac{\partial}{\partial y} + \dfrac{\partial}{\partial z}\otimes \dfrac{\partial}{\partial z}
$$       
and depending on the behaviour of the potential function this symmetric tensor may have different properties.  It could define a metric tensor and one could introduce a relative Laplace-Beltrami operator.

We go back now to build the Hilbert space  of differential forms starting from a Riemannian manifold $M$ equipped with a metric tensor $g$. The contravariant form $\tilde g$ of the metric tensor allows to define the following scalar product between differential 1-forms $\alpha, \, \beta \in \Lambda^1(M)$:
\begin{equation}
\left( \alpha | \beta \right) = \tilde{g}(\alpha,\beta). 
\end{equation}
This scalar product is extended to higher degree differential forms. Forms of different degrees are declared to be orthogonal.  Let us now consider two differential forms of degree k, written in terms of one forms as $\alpha = \alpha_1 \wedge \cdots \wedge \alpha_k$ and $\beta = \beta_1 \wedge \cdots \wedge \beta_k$; their scalar product is defined as
\begin{equation}
\left( \alpha| \beta \right) = {\rm det} \left( \alpha_j | \beta_k \right)
\end{equation}  
where $\alpha_j, \, \beta_j  \in \Lambda^1(M)$, and the determinant is intended with respect to the matrix indices $j,k$. 
One can rewrite the previous scalar product in terms of the Hodge dual operator $$
\ast \: : \: \Lambda^k(M) \; \rightarrow \; \Lambda^{m-k}(M)
$$
as 
\begin{equation}
\alpha \wedge \ast \beta = \left( \alpha | \beta \right) \Omega
\end{equation}    
where $\Omega$ is the metric volume form which in a chart can be written as 
$$
\Omega = \sqrt{|det(g)|} dx^1 \wedge \cdots \wedge dx^m
$$
on a local chart for $M$.
The Hilbert space of square integrable differential forms 
$\mathcal{L}_{\Lambda}^2(M,d\mu)$ is given by means of the following product
\begin{equation}
\left\langle \alpha | \beta \right\rangle = \int_M \left( \alpha, \beta \right) d\mu
\label{scalprod}
\end{equation}
where $d\mu$ is the measure associated to the volume form $\Omega$.

Let us consider $M$ as a manifold whose dimension is $n$. One defines the codifferential operator $\delta : \Lambda^k(M) \, \rightarrow \, \Lambda^{k-1}(M)$  as
$$
\delta= (-1)^{n(k-1)+1}\ast d \ast
$$
if the metric is Riemannian, or
$$
\delta = (-1)^{n(k-1)}\ast d \ast
$$
if the metric is Lorentzian. When the manifold has not a boundary the codifferential is the adjoint operator of the exterior derivative with respect to the scalar product \eqref{scalprod}, that is 
$$
(\alpha\,|\,d\beta)\,=\,(\delta\alpha\,|\,\beta).
$$
It is also easy to see that $\delta^2=0$, and that the  action of the Laplace-Beltrami operator can be extended to differential forms according to the formula
$$
\Delta = d \delta + \delta d = (d+\delta)^2
$$
This expression is very interesting as we will see in the following because it allows to define immediately a ``square-root" of the Laplace-Beltrami operator obtaining a Dirac-type operator written in terms of scalar differential operators instead of matrix-valued differential operators.

Another advantage related to the introduction of differential forms in quantum mechanics consists in the fact that differential forms contain information about the topology of the carrier space in a more direct way. For instance cohomology theory allows to extract information about a manifold just looking at some subspaces of differential forms.\newline
In the rest of the paper we will show some applications of this formalism related to vector-valued harmonics and to Dirac-type operators.

\subsection{Vector-valued harmonics as differential forms}
In this section we show -- via an example --  that an Hilbert space of differential forms carries interesting representations for some Lie algebras. 
 According to representation theory, unitary  representations of a compact Lie group $G$
 on a separable Hilbert space $\mathcal{H}$ can be written as a suitable direct sum of  finite dimensional irreducible ones, and $\mathcal{H}$ itself can be written as a suitable direct sum of spaces upon which the representations are irreducible. Irreducibility subspaces are labelled by the eigenvalues of the Casimir operators for the Lie algebra $\mathfrak{g}$ corresponding to $G$. 
  For instance the Hilbert space $\mathcal{L}^2({\rm S}^2,\sin \theta d\theta \varphi)$ of square-integrable functions on the sphere can be decomposed into finite dimensional vector spaces of dimension $d=2l+1$, for $l=0,1,\cdots$, on which the  left-action of the rotation group is represented in terms of matrices. A basis for each of these subspaces is given by the spherical harmonics, and the index $l$ characterizes the spectrum of the Casimir of the Lie algebra of the rotation group.

A similar construction can be also realized for the space of differential forms. As a final result we will show that the space of differential one-forms can be used as vector space for representations of the algebra $\mathfrak{su}(2)$ with eigenvalue of the Casimir operator both integer and half-integer. 
Let us consider the cotangent bundle ${\rm T^*S}^2$ of the sphere ${\rm S}^2$. It is known that this space is not parallelizable, that is it does not admit a globally defined differential calculus, and then an exterior algebra, on it. 
The sphere ${\rm S}^2$ is nevertheless an homogeneous space, i.e. it is the base space of the Hopf principal bundle $\pi \: : \: {\rm S}^3 \rightarrow {\rm S}^2$ with gauge group ${\rm U}(1)$. This is the principal bundle also considered in the analysis of the  the electron-monopole system. We consider as metric the Killing-Cartan metric written in terms of left invariant differential forms as 
\begin{equation}
g=\theta^1\otimes \theta^1+\theta^2\otimes \theta^2+\theta^3\otimes \theta^3
\end{equation}
The exterior algebra over ${\rm S}^2$ can be written as a suitable sub algebra of the exterior algebra $\Lambda({\rm S}^3)$, namely those elements $\alpha\,\in\,\Lambda({\rm S}^3)$ fulfilling the conditions 
\begin{equation}
\begin{split}
L_{X_3} \alpha = 0 \\
i_{X_3}\alpha = 0
\end{split}
\end{equation}       
where $X_3$ is the left invariant field generating the $U(1)$ action on ${\rm S}^3$. Differential forms satisfying this conditions are of the kind
\begin{equation}
\alpha = \alpha_+ \theta^+ + \alpha_-\theta^-
\end{equation} 
where $\theta^{\pm}=\frac{1}{\sqrt{2}}\left( \theta^1 \mp i \theta^2 \right)$ and the coefficients obey to the following relations
\begin{equation}
\begin{split}
X_3\alpha_+ = -i\alpha_+ \\
X_3 \alpha_-= i \alpha_- 
\label{inv}
\end{split}
\end{equation}
These differential forms are a Hilbert subspace of the Hilbert space of the Hilbert space of square integrable differential forms on ${\rm S}^3$ because the coefficients form a Hilbert subspace of the space $\mathcal{L}^2({\rm S}^3,d\nu)$ where $d\nu$ is the Haar measure on ${\rm S}^3$. Indeed, as the fibre is a compact space, functions which are square-integrable on ${\rm S}^3$ are still square-integrable on the base manifold.

It is possible to define  on this subspace a representation of the algebra $\mathfrak{su}(2)$ generated by the right-invariant vector fields $L_a$, which are the infinitesimal generators of the left action of the group ${\rm SU}(2)$ on ${\rm S}^3$. Therefore one has to find a set of common eigenforms $\alpha\in \Lambda({\rm S}^3)$ of the operators $\left\lbrace L^2, L_z \right\rbrace $ by solving the following equations
\begin{equation}
\begin{split}
L_z \alpha = i m \alpha \\
L^2 \alpha = -j(j+1)\alpha
\end{split}
\end{equation} 
As $\theta^a$ are left-invariant differential forms they are in the kernel of the right invariant vector fields. Therefore, taking into account also the condition $\eqref{inv}$, an orthonormal basis for a $j=1$ representation is the following 
\begin{equation}
\begin{split}
&\alpha_1=i\dfrac{1}{\sqrt{\pi}}\sqrt{\dfrac{3}{8\pi}}\left[ v^2(\bar{v}d\bar{u}-\bar{u}d\bar{v}) + \bar{u}^2(udv-vdu) \right] \\
&\alpha_0=i\dfrac{1}{\sqrt{\pi}}\sqrt{\dfrac{3}{4\pi}}\left[ -vu(\bar{v}d\bar{u}-\bar{u}d\bar{v}) + \bar{v}\bar{u}(udv-vdu) \right] \\
&\alpha_{-1}=i\dfrac{1}{\sqrt{\pi}}\sqrt{\dfrac{3}{8\pi}}\left[ u^2(\bar{v}d\bar{u}-\bar{u}d\bar{v}) + \bar{v}^2(udv-vdu) \right]
\label{j=1}
\end{split}
\end{equation}
Since  invariant differential forms are a module over the algebra $\mathcal{F}(S^2)\,=\,{\rm Kern}\,X_3\,\subset\,\mathcal{F}({\rm S}^3$, one can build combinations of differential forms and spherical harmonics in $\mathcal{F}({\rm S}^2)$ with Clebsch-Gordan coefficients. In this way one obtains higher order integer representations. These differential forms are in correspondence with the vector valued harmonics defined in $\cite{Marmo}$.

A step forward can be done by replacing  the invariance condition in $\eqref{inv}$ with the equivariance condition $\eqref{eq2}$. In fact if one chooses the subspace of equivariant differential forms with eigenvalue $n=\frac{1}{2}$ it is possible to construct a representation with $j=\dfrac{1}{2}$. The condition $\eqref{inv}$ is replaced by the following      
\begin{equation}
\begin{split}
&X_3\alpha_+ = -\dfrac{i}{2}\alpha_+ \\
&X_3 \alpha_-= i\dfrac{3}{2} \alpha_- 
\label{equiv}
\end{split}
\end{equation}
Since we are interested in a $j=\dfrac{1}{2}$ representation of the rotation algebra, we consider only differential forms of the kind $\tilde{\alpha}=\alpha_+ \theta^+$. An orthonormal basis for this representation is given by the following eigenforms of $\left\lbrace L^2, L_z \right\rbrace $
\begin{equation}
\begin{split}
&\alpha_{\frac{1}{2}}=\dfrac{1}{2\pi}v\theta^+ \\
&\alpha_{-\frac{1}{2}}=-\dfrac{1}{2\pi}u\theta^+
\label{basis}
\end{split}
\end{equation}
Also equivariant differential forms are a module over the ring $\mathcal{F}({\rm S}^2)$. Therefore it is possible to build higher order half-integer representations taking combinations of the differential forms $\eqref{basis}$ and spherical harmonics in $\mathcal{F}(S^2)$ with Clebsch-Gordan matrix elements as coefficients. As an example of this construction one can build the representation which is the product of the representation with $j=\frac{1}{2}$ and the spherical harmonics in $\mathcal{F}(S^2)$ with $j=1$. Properly using the  right Clebsch-Gordan coefficients one can write down the basis for the representations with $j=\frac{3}{2}$ and $j=\frac{1}{2}$. The final results are
\begin{equation}
\begin{split}
&\alpha_{\frac{3}{2}}=Y_1^1\alpha_{\frac{1}{2}} \\
&\alpha_{\frac{1}{2}}=\sqrt{\frac{1}{3}} Y_1^1\alpha_{\frac{1}{2}}+\sqrt{\frac{2}{3}}Y_1^0\alpha_{-\frac{1}{2}} \\
&\alpha_{-\frac{1}{2}}=\sqrt{\frac{1}{3}} Y_1^{-1}\alpha_{\frac{1}{2}}+\sqrt{\frac{2}{3}}Y_1^0\alpha_{-\frac{1}{2}}\\
&\alpha_{-\frac{3}{2}}=Y_1^{-1}\alpha_{-\frac{1}{2}}
\end{split}
\end{equation}
for $j=\dfrac{3}{2}$ and
\begin{equation}
\begin{split}
&\alpha_{\frac{1}{2}}=\sqrt{\frac{2}{3}}Y_1^1\alpha_{\frac{1}{2}}-\sqrt{\frac{1}{3}}Y_1^0\alpha_{-\frac{1}{2}}\\
&\alpha_{-\frac{1}{2}}=-\sqrt{\frac{2}{3}}Y_1^{-1}\alpha_{\frac{1}{2}}+\sqrt{\frac{1}{3}}Y_1^0\alpha_{-\frac{1}{2}}
\end{split}
\end{equation}
for $j=\dfrac{1}{2}$.\newline  
In summary in this section  we have briefly shown that the Hilbert space of differential forms can be used in representation theory to write down  vectorial representations of the algebra $\mathfrak{su}(2)$ with both integer and half-integer eigenvalues of the Casimir operator. This possibility can be useful especially in relation to gravitational problems. In fact this kind of construction can be repeated for other homogeneous group, giving rise to different kinds of tensor harmonics. These objects are useful to decompose tensors on a basis which respects the action of some transformation group.

\subsection{Algebraic spinors and Dirac-type operators} 
Another application regards the possibility of writing, on a manifold $M$ equipped with a metric tensor $g$,  Dirac-type operators in terms of scalar -- which means invariant under the action of an element in ${\rm Diff}\,(M)$, the group of diffeomorphisms on $M$ -- differential operators on a manifold. On such a manifold $(M, g) $  the Laplace-Beltrami operator can be written, as we already mentioned,  in terms of differential and codifferential as follows
$$
\Delta = (d + \delta)^2.
$$
Therefore one can write  a Dirac operator as a  (up to a constant factor) square-root  
$$
D= ( d - \delta )
$$           
acting upon the whole exterior algebra $\Lambda(M)$. This idea goes back to K\"{a}hler who wrote a representation of a Clifford product (the so called inner, or $\vee$-product) on $(\Lambda(M), \wedge)$\cite{Kahler}.
The inner calculus is defined by introducing a Clifford product on $\Lambda(M)$. If the metric tensor $g$ on $M$ has the local coordinate expression $g\,=\,g_{ab}\dd x^a\otimes\dd x^b$ or equivalently $g\,=\,g^{ab}\del_a\otimes\del_b$ with $g^{ab}g_{bc}\,=\,\delta^a_c$, one has
\beq
\label{clm}
 \phi\vee\phi^{\prime}\,=\,\sum_s\,\frac{(-1)^{\tiny{\left(\begin{array}{c}s \\ 2 \end{array}\right)}}}{s!}g^{{a_1}{b_1}}\,\cdots\,g^{{a_s}{b_s}}(\gamma^s\{i_{a_1}\,\cdots \,i_{a_s}\,\phi\})\wedge\{i_{b_1}\,\cdots\,i_{b_s}\,\phi^{\prime}\}, 
\eeq
where $\phi,\,\phi^{\prime}$ are elements in $\Lambda(M)$, one has  $\gamma(\phi)\,=\,(-1)^k\phi$ for $\phi\,\in\,\Lambda^k(M)$ ($\gamma$ is the degree operator) and $i_{a}\,=\,i_{\del_a}$ is the contraction operator. One clearly has
\begin{align}
\dd x^a\vee\dd x^b&=\,\dd x^a\wedge\dd x^b\,+\,g^{ab}, \nn \\
\dd x^a\vee\dd x^b\,+\,\dd x^b\vee\dd x^a&=\,2g^{ab}.
\label{esv}
\end{align}
On a local chart on $M$,  the Dirac operator is defined by 
\begin{equation}
D=\sum_{a=1}^m dx^a \vee \nabla_a 
\end{equation} 
where $\left\lbrace dx^a \right\rbrace$ is a local basis of the cotangent bundle ${\rm T^*M}$ and $\nabla_a=\nabla_{\frac{\partial}{\partial x_a}}$ is the Levi-Civita covariant derivative. When  acting upon a differential form, it gives 
\begin{equation}
D\phi = i(d+(-1)^{N(k-1)}\ast d \ast)\phi = (d- \delta )\phi.
\label{Dirac}
\end{equation}
The paper \cite{Graf} shows how it is possible to decompose the left action of the Clifford algebra $(\Lambda(M), \vee)$ on itself. Irreducible modules $I_j\,\subset\,\Lambda(M)$ correspond to ranges of projectors $P_j\,\in\,\Lambda(M)$, i.e. $P_j\vee P_j\,=\,P_j$. Elements in $I_j$ are called algebraic spinors since they carry an action of the Spin group corresponding to the metric tensor $g$. The Dirac operator $D$ \eqref{Dirac} turns out to be meaningful when restricted to $I_j$ if and only if $P_j\vee\nabla_aP_j\,=\,0$. 

We want now to show how the Dirac-Pauli operator on $\R^3$ can be written in terms of this formalism. We consider  
$$
(\R^3,  g\,=\,d x\otimes d x\,+\,d y\otimes d y\,+\,d z\otimes d z)
$$ 
whose corresponding Hodge duality reads
\begin{align}
\star(1)\,=\,\tau\,=\,\dd x\wedge\dd y\wedge\dd z &\qquad\qquad\star(\tau)\,=\,1 \nonumber \\
\star\dd x\,=\,\dd y\wedge\dd z &\qquad\qquad \star(\dd y\wedge\dd z)\,=\,\dd x, \nonumber \\
\star\dd y\,=\,\dd z\wedge\dd x &\qquad\qquad \star(\dd z\wedge\dd x)\,=\,\dd y, \nonumber \\
\star\dd z\,=\,\dd x\wedge\dd y &\qquad\qquad \star(\dd x\wedge\dd y)\,=\,\dd z. 
\label{hs3}
\end{align}
We focus our attention to a class of real (which means with real coefficients) solutions for the equation $P\vee P\,=\,P$ in $\Lambda(\R^3)$, given by 
\begin{equation}
\label{p3di}
P(\xi)\,=\,\frac{1}{2}\,+\,\rho\,\dd x\,+\,\xi\,\dd x\wedge\dd y
\end{equation}
with $4\rho^2\,=\,4\xi^2+1$ and $\rho\,>\,0, \,\xi\,\geq\,0$. Projectors are then labelled by $\xi$.
The range of the action of the projector $P(\xi)$ gives the   left ideal $I_{\xi}$, which is then a set of spinors. A basis for $I_{\xi}$ turns out to be:
\begin{align}
&\psi_1\,=\,1\,+\,2\rho\,\dd x\,+\,2\xi\,\dd x\wedge\dd y, \nonumber \\
&\psi_2\,=\,\dd y\,-\,2\xi\,\dd x\,-\,2\rho\,\dd x\wedge\dd y, \nonumber \\
&\psi_3\,=\,\dd y\wedge\dd z\,+\,2\xi\,\dd z\wedge\dd x\,+\,2\rho\,\tau, \nonumber \\
&\psi_4\,=\,\dd z\,+\,2\rho\,\dd z\wedge\,\dd x\,+\,2\xi\,\tau \label{baspi3}.
\end{align} 
One can write 
$I_{\xi}\,\ni\,\psi\,=\,\sum_af_a\psi_a$ with $f_a\,\in\,{\cal F}(\R^3)$. One gets 
\begin{align}
&\dd x\,\vee\psi\,=\,\left(\begin{array}{cccc} 2\rho & -2\xi & 0 & 0 \\ 2\xi & -2\rho & 0 & 0 \\ 0 & 0 & 2\rho & 2 \xi \\ 0 & 0 & -2\xi & -2\rho \end{array}\right)
\left(\begin{array}{c} f_1 \\ f_2 \\ f_3 \\ f_4 \end{array}\right), \label{ut3} \\
&\dd y\,\vee\psi\,=\,\left(\begin{array}{cccc} 0 & 1 & 0 & 0 \\ 1 & 0 & 0 & 0 \\ 0 & 0 & 0 & 1 \\ 0 & 0 & 1 & 0 \end{array}\right)
\left(\begin{array}{c} f_1 \\ f_2 \\ f_3 \\ f_4 \end{array}\right), \label{ut4} \\
&\dd z\,\vee\psi\,=\,\left(\begin{array}{cccc} 0 & 0 & 0 & 1 \\ 0 & 0 & -1 & 0 \\ 0 & -1 & 0 & 0 \\ 1 & 0 & 0 & 0 \end{array}\right)
\left(\begin{array}{c} f_1 \\ f_2 \\ f_3 \\ f_4 \end{array}\right). \label{clact}
\end{align}
The action of the Dirac operator turns out to be well defined on $I_{\xi}$.  
 Such an action can be given the following matrix form, if one considers a spinors $I_{\xi}\,\ni\,\psi\,=\,\sum_af_a\psi_a$ with $f_a\,\in\,{\cal F}(\R^3)$:
\begin{equation}
\label{d3a}
-i\mathcal{D}\left(\begin{array}{c} f_1 \\ f_2 \\ f_3 \\ f_4 \end{array}\right)\,=\,\left(\begin{array}{cccc} 2\rho\del_x & \del_y-2\xi\del_x & 0 & \del_z \\ 
\del_y+2\xi\del_x & -2\rho\del_x & -\del_z & 0 \\ 0 & -\del_z & 2\rho\del_x & 2\xi\del_x+\del_y \\ \del_z & 0 & \del_y-2\xi\del_x & -2\rho\del_x 
\end{array}\right) \left(\begin{array}{c} f_1 \\ f_2 \\ f_3 \\ f_4 \end{array}\right).
\end{equation}
The K\"ahler Dirac operator does not in general coincide with the spin manifold Dirac operator (see \cite{Lawson}). When the metric tensor gives a flat Levi Civita connection, then the two operators may coincide. We are now going to describe how, starting from \eqref{d3a}, one can write down the Pauli Dirac operator, which is
\begin{equation}
\label{spMD}
\tilde{D}\phi\,=\,\sigma_a\otimes\del_a\phi
\end{equation}
where $\phi\in\,\mathcal{F}(\R^3)\otimes\C^2$ is a spinor field, i.e. a section of the vector bundle $\R^3\times\C^2$. 

The action \eqref{d3a} of the Dirac operator is irreducible on $I_{\xi}$, and seems to be quite far from the action of the spin manifold Dirac operator $\tilde{D}$ in \eqref{spMD}. A possible path bringing the action of $D$ closer to that of $\tilde{D}$ starts by noticing that the volume element $\tau\,=\,dx\wedge dy\wedge dz$  satisfies the identity $\tau\vee\tau\,=\,-1$. We  define then by $J$ the matrix acting on $I_{\xi}$ that represents the volume form $\tau$, i.e. we define  
\begin{equation}
\label{Jc}
J\,=\,\dd x\vee\dd y\vee \dd z\,=\,\left(\begin{array}{cccc} 0 & 0 &  -2\rho & -2\xi \\ 0 & 0 & -2\xi & -2\rho \\ 2\rho & -2\xi & 0 & 0 \\ -2\xi & 2\rho & 0 & 0 \end{array}\right)
\end{equation}
with $J\vee J\,=\,-1$ with respect to the basis $\{\psi_1, \ldots, \psi_4\}$. This is equivalent to  write 
\begin{align}
&J\vee\psi_1\,=\,2\rho\psi_3 \, -\,2\xi\psi_4, \nonumber \\
&J\vee\psi_2\,=\,-2\xi\psi_3 \, +\,2\rho\psi_4, \nonumber \\
&J\vee\psi_3\,=\,-2\rho\psi_1 \, -\,2\xi\psi_2, \nonumber \\
&J\vee\psi_4\,=\,-2\xi\psi_1 \, -\,2\rho\psi_2. 
\label{jbas}
\end{align}
The endomorphism $J$ defines a complex structure over the four dimensional ideal $I_{\xi}$. 
Along  the basis $B\,=\,\{\psi_{1}, \psi_2, J\vee\psi_1, J\vee\psi_2\}$   for $I_{\xi}$ the endomorphism  $J$ has the canonical form, 
so upon identifying the action  $J\,\vee$ with the multiplication by \emph{an} imaginary unit $i$, the left ideal $I_{\xi}$ is spanned by complex valued coefficients along the real basis elements $\{\psi_1,\psi_2\}$. 
Since one proves that 
\beq 
\label{JvD}
\D(f_a\,J\,\vee\,\psi_a)\,=\,J\,\vee\,(\D\,(f_a\,\psi_a)),
\eeq 
the action of the Dirac operator is consistently reduced to $\mathcal{F}(\R^3)\otimes_{\R}\C\otimes_{\R}\{\psi_1, \psi_2\}$, which is a space of two-dimensional complex spinors over $\R^2$:
\beq
\label{dbd}
\D\left(\begin{array}{c} f_1+ih_3 \\ f_2 +ih_4\end{array}\right)\,=\,
\left(\begin{array}{cccc} 2\rho\del_x+2i\xi\del_z & \del_y-2\xi\del_x-2i\rho\del_z &  \\ \del_y+2\xi\del_x +2i\rho\del_z & -2\rho\del_x-2i\xi\del_z  
\end{array}\right)\,
\left(\begin{array}{c} f_1+ih_3 \\ f_2 +ih_4 \end{array}\right).
\eeq 
For $\xi\,=\,0$ and $\rho\,=\,1/2$ it turns out to be equivalent to the spin manifold Dirac operator \eqref{spMD}. 
The action of this Dirac operator can be written as 
\beq 
\D\psi\,=\,\dd x^a\otimes\del_a\psi
\label{ditir}
\eeq
 after defining 
an algebra  map -- over $\R$ -- $(\Lambda(\R^3), \vee)\,\to\,{\rm M}^2(\C)$ via
\beq 
\label{clco}
\dd x\,\mapsto\,\left(\begin{array}{cc}2\rho & -2\xi \\ 2\xi & -2\rho \end{array}\right),\qquad\dd y\,\mapsto\,\left(\begin{array}{cc} 0 & 1 \\ 1 & 0 \end{array}\right), \qquad\dd z\,\mapsto\,i\,\left(\begin{array}{cc}2\xi & -2\rho \\ 2\rho & -2\xi \end{array}\right).
\eeq 
that turns out to be an irreducible representation of $(\Lambda(\R^3), \vee)$ on $\C^2$.

\subsection{The Dirac-K\"ahler operator on $\R^3$ upon complexification}
\label{sscom}

On the space  $\Lambda(\R^3)\otimes_{\R}\C$ 
are both the wedge and the Clifford products  well defined, so we may define the Clifford algebra
$(\Lambda(\R^3)\otimes_{\R}\C, \vee)$. The  element 
\beq
\label{pc3}
P\,=\,\frac{1}{4}(1+\dd z+i\dd x\wedge\dd y+i\dd x\wedge\dd y\wedge\dd z)
\eeq
is an idempotent with respect to the Clifford product and its range $I_P$ is a two dimensional left ideal for the Clifford algebra $(\Lambda(\R^3)\otimes_{\R}\C, \vee)$ whose basis is given by
\begin{align}
\psi_1\,=\,&1+\dd z+i\dd x\wedge\dd y+i\dd x\wedge\dd y\wedge\dd z, \nn \\
\psi_2\,=\,&\dd x+i\dd y+i\dd y\wedge\dd z+\dd x\wedge\dd z.
\label{ba3c}
\end{align}
From the identities 
\begin{align}
&\dd x\vee\psi_1\,=\,\psi_2, \nn \\ 
&\dd y\vee\psi_1\,=\,-i\psi_2, \nn \\
&\dd z\vee\psi_1\,=\psi_1, \qquad \dd z\vee\psi_2\,=\,-\psi_2,
\label{id3ra}
\end{align}
it is straightforward to see that the action of the generators of the Clifford algebra upon $I_P$ is given as a matrix products by the Pauli matrices, i.e. 
\beq
\label{3pau}
\dd x\,\vee\,\mapsto\,\left(\begin{array}{cc} 0 & 1 \\ 1 & 0 \end{array}\right), \qquad
\dd y\,\vee\,\mapsto\,\left(\begin{array}{cc} 0 & i \\ -i & 0 \end{array}\right), \qquad
\dd z\,\vee\,\mapsto\,\left(\begin{array}{cc} 1 & 0 \\ 0 & -1 \end{array}\right).
\eeq
For the spinor space one has $I_P\,=\,\mathcal{F}(\R^3)\,\otimes_{\R}\C^2$, that is spinors are two components complex valued functions defined on $\R^3$. Upon such spinors $\psi\,\in\,I_P$, the action of the Dirac operator can be represented by  
\beq
\D\,=\,\dd x^a\,\vee\,\nabla_a\,=\,\sigma^{a}\otimes\del_a.
\label{di3com}
\eeq

\section{Conclusions}
In this paper we have revisited the approach to electron-monopole proposed by Balachandran et al, obtaining it by reduction of a geodesical motion on a bigger space. Considering the possibility of writing Hamiltonian operators in terms of Laplace-Beltrami operator, we have proposed to introduce the generalized Hilbert space of square integrable differential forms. 

The usual transition from flat space-time to Lorentzian manifolds considers the flat space as the tangent space at each given point of the manifold. In this generalization one encounters the Bochner calculus and the subsequent elaboration by Lichnerowicz.
Our idea is to generalize the theory from $\R^4$ by considering it as a Lie group, therefore the simplest generalization would be to go from an Abelian vector group to a non-Abelian one. This approach has the advantage that we can always work with parallelizable manifolds,  and it can be applied also to homogeneous spaces when a reduction with respect to a closed subgroup is considered.  We hope to be able to tackle also the situation of manifolds with boundaries when a quotient procedure of homogeneous spaces with respect to discrete transformations is conceived.
Having this in mind,in this paper, 
we have applied this idea to two situations. The first case is related to the theory of group action on homogeneous manifold: we have written tensor harmonics for the rotation group in terms of differential forms. The second application regards spin geometry: it is possible, in fact, to write the Dirac operator as the square root of the Laplace-Beltrami operator, using the exterior derivative $d$ and the codifferential $d^{\dagger}$.


\begin{thebibliography}{}
\bibitem{Poincare} H. Poincaré, Compt. Rend. Acad. Sci. Paris, 1896
\bibitem{Dirac}P.A.M. Dirac, Proc. Roy. Soc. A133 (1931) 60
\bibitem{Wu-Yang}T.T. Wu, C.N.Yang, Nucl. Phys. B107 (1976) 365
\bibitem{balga} A.P. Balachandran, G. Marmo, A. Simoni, G. Sparano, Int. J. Mod. Phys. A07  (1992) 1641
\bibitem{Bal1} A.P. Balachandran, G. Marmo, B.-S. Skagerstam, A. Stern , Nucl. Phys. B162 (1980)385
\bibitem{Bal2}A.P. Balachandran, G. Marmo, B.-S. Skagerstam, A. Stern \emph{Gauge Symmetries and Fiber Bundles}, Lect. Notes in Phys. 188, Springer, Berlin, 1983
\bibitem{Bal3} A.P. Balachandran,  G. Marmo, B.-S. Skagerstam, A. Stern, \emph{Classical topology and Quantum States}, World Scientific, Singapore, 1991
\bibitem{Hertz} H.R. Hertz, \emph{The Principles of Mechanics Presented in a New Form}, Macmillan, London, 1899 (English translation of Die Prinzipien der Mechanik in neuem Zusammenhange dargestellt, Leipzig, posthumously published in 1894)  
\bibitem{kaluza-Klein} T. Kaluza, Sitz.Prens.Akad.Wiss. (1921), 966\newline
O. Klein, Z.Phys. 37 (1926), 895
\bibitem{Duval} C. Duval, G. Burdet, H.P. K\"{u}nzle, M.Perrin, Phys. Rev. D31 (1985) 1841
\bibitem{Lizzi} F. Lizzi, G. Marmo, G. Sparano, A.M. Vinogradov, J. Geom. Phys. 14-3 (1994) 211
\bibitem{Grabowski} J. Grabowski, K. Grabowska, P. Urbanski, J. Phys. A41, 14 (2008) 5204
\bibitem{Dirac2} P.A.M. Dirac, \textit{Lectures on quantum mechanics}, Belfer Graduate School of Science Monographs series 2 (Yeshiva University, New York , 1964)
\bibitem{Marmo3}G. Marmo, C. Rubano, \textit{Particle Dynamics on Fibre bundles}, Bibliopolis, Napoli, 1988
\bibitem{Bergmann} A. Einstein, P. Bergmann, Ann. of Math. \textbf{39}(1938), 683
\bibitem{Dray} T. Dray, J. Mat. Phys. 27 (1981) 781
\bibitem{Schwinger rep} S. Chaturvedi, G. Marmo, N. Mukunda, R. Simon, A. Zampini, Rev. Mod. Phys. 18 (2006) 881
\bibitem{Carinena} J.F. Cari\~{n}ena, A. Ibort, G. Marmo, G. Morandi, \textit{Geometry from dynamics, classical and quantum}, Springer, Dordrecht, 2015
\bibitem{Marmo} G. Marmo, \textit{Differential forms and Electrodynamics}, (1986) (Lectures at Shanxi University)
\bibitem{Kahler} E. K\"{a}hler, Rendiconti di Matematica vol. 21 (1962) 425
\bibitem{Graf}W. Graf, Ann. I.H.P., section A, 29 (1978) 85
\bibitem{Lawson}H.B. Lawson, M.L. Michelson, \textit{Spin Geometry},  Princeton University Press, Princeton, New Jersey, 1989
\end{thebibliography}
\end{document}